\documentclass[preprint,12pt]{elsarticle}

\usepackage{amssymb}

\usepackage{subcaption}

\usepackage{footmisc}

\usepackage{tikz}
\usetikzlibrary{shapes.misc,shadows}
\usetikzlibrary{shapes,snakes}
\usetikzlibrary{automata,positioning}

\usepackage{hyperref}
\usepackage{amsmath}

\usepackage{listings}
\usepackage{xcolor} 

\definecolor{mygreen}{rgb}{0,0.6,0}
\definecolor{mygray}{rgb}{0.5,0.5,0.5}
\definecolor{mynavyblue}{rgb}{0.1,0.1,0.4}

\newcounter{bla}

\journal{Computer Physics Communications}

\begin{document}

\begin{frontmatter}



\title{Analytic calculator for determination of $\gamma$-ray angular distribution coefficients and tensors in aligned and partially-aligned nuclei}


\author[a,b]{A.M.~Hurst\corref{author}}
\cortext[author] {Corresponding author.\\\textit{E-mail address:} amhurst@berkeley.edu}
\address[a]{Department of Nuclear Engineering, University of California, Berkeley, California, 94720, USA}
\address[b]{Nuclear Science Division, Lawrence Berkeley National Laboratory, Berkeley, California, 94720, USA}

\author[b]{D.A.~Matters}

\author[c]{T.~Kawano}
\address[c]{Theoretical Division, Los Alamos National Laboratory, Los Alamos, NM 87545, USA}

\begin{abstract}

A program has been developed to calculate a complete set of $\gamma$-ray angular distribution coefficients and statistical tensors in maximally- and partially-aligned nuclei.  For practical nuclear structure and reaction purposes, there is no imposed constraint on any arguments that are likely to arise in the determination of these quantities.  The program can also be used as a stand-alone vector-coupling calculator for the exact evaluation of Clebsch-Gordan and Racah coefficients, the closely-related Wigner 3-$j$, 6-$j$, and 9-$j$ symbols, as well as Gaunt coefficients.  These quantities, which frequently arise in quantum mechanical applications involving angular momentum coupling and recoupling schemes, provide the underlying machinery in angular distribution calculations.





\end{abstract}

\begin{keyword}
$\gamma$-ray angular distributions; statistical tensors; angular momentum; Clebsch-Gordan coefficients; Wigner coefficients; 3-$j$ symbol; Racah coefficients; 6-$j$ symbol; 9-$j$ symbol; Gaunt coefficients; nuclear structure

\end{keyword}

\end{frontmatter}



{\bf PROGRAM SUMMARY}

\begin{small}
\noindent
{\em Program Title:} \texttt{PyGammaRAD}: Python project for Gamma-Ray Angular Distributions\\
{\em CPC Library link to program files:} (to be added by Technical Editor) \\
{\em Developer's repository link:} \url{https://pypi.org/project/PyGammaRAD/} and \url{https://github.com/AaronMHurst/PyGammaRAD} \\
{\em Code Ocean capsule:} (to be added by Technical Editor)\\
{\em Licensing provisions:} MIT license \\
{\em Programming language:} {\tt Python} \\
{\em Nature of problem:} Calculations based on angular momentum coupling schemes needed for the determination of $\gamma$-ray angular distribution coefficients and statistical tensors in maximally- and partially-aligned nuclei, in addition to other quantum mechanical applications that require the exact evaluation of angular momentum vector-coupling coefficients.\\
{\em Solution method:} Direct evaluation of sum formulae using conventional numerical computing methods to calculate angular momentum coupling coefficients and symbols, and the angular distribution coefficients and statistical tensors derived from these quantities.\\
   \\

\end{small}

\section{\label{sec:intro}Introduction}

The angular distribution of $\gamma$ rays emitted from oriented nuclear states carries rich spectroscopic information, encoding the multipolarities $XL$, where $X$ denotes the electric (E) or magnetic (M) character and $L$ the multipole order, of $\gamma$-ray transitions together with the associated spin $J$ and parity $\pi$ assignments of the nuclear levels involved. Many important parameters associated with angular distributions routinely used in nuclear physics data analysis are only available in limited tabular forms and there is a need to have libraries that support direct calculation to enable future automated analysis techniques and applications.  However, before discussing the utility and formalism of $\gamma$-ray angular distributions, it is worthwhile to establish precise definitions, since the literature is not always unambiguous on this point.

A nuclear state of spin $J$ possesses $2J+1$ magnetic substates, each characterized by a quantum number $m$ taking values from $-J$ to $+J$, defined with respect to a chosen quantization axis. When the populations of all magnetic substates are equal, the nucleus is said to be \textit{unoriented}, and the resulting $\gamma$-ray angular distribution is perfectly isotropic, e.g., an unoriented $J=2 \,\hbar$ state has five associated magnetic substates each with equal population: $P(-2) = P(-1) = P(0) = P(+1) = P(+2) = \frac{1}{5}$.  Any departure from this uniform population constitutes nuclear \textit{orientation}.  There are two classes of oriented nuclear states: \textit{aligned} and \textit{polarized}.  Both classes of orientation lead to anisotropies in the observed $\gamma$-ray angular distribution depending on the degree of alignment or polarization.  A nucleus is said to be \textit{aligned} if the population distribution is symmetric about $m=0$, thus, our alignment condition is defined by $P(-m) = P(m)$.  In the example of an aligned $J=2 \,\hbar$ state, this naturally implies $P(-2)=P(+2)$ and $P(-1)=P(+1)$, while the symmetry about $m=0$ necessarily ensures that $P(|2|) < P(|1|) < P(0)$.  In \textit{polarized} nuclear states this symmetry is broken and $P(-m) \neq P(m)$.  Consequently, for a polarized $J=2 \,\hbar$ state, to use the same example, this implies $P(-2) \neq P(+2)$ and $P(-1) \neq P(+1)$.

In reactions involving a beam of particles incident upon a target nucleus, such as compound-nucleus reactions \cite{bohr:36}, the beam direction naturally defines the quantization axis, often referred to as the $z$ axis.  Here, the reacting system possesses a reflection symmetry about the plane containing the beam axis for an unpolarized beam incident on an unpolarized target.  This guarantees that the residual nuclear state is aligned but not polarized; a consequence of the axial-reflection symmetry of the reacting system \cite{litherland:61}. Aligned states are further classified into two sub-classes depending on the degree of alignment: \textit{maximum} \textit{alignment} and \textit{partial alignment}.  When the alignment is modeled as a Gaussian distribution of magnetic-substate populations \cite{yamazaki:67, mateosian:74} characterized by a width parameter $\sigma$, the degree of alignment is controlled by the ratio $\sigma/J$. In the limit of small $\sigma/J$ (i.e., as $\sigma/J \rightarrow 0$) the Gaussian is narrow and population concentrates near $m = 0$, corresponding to the nuclear spin vector aligned predominantly perpendicular to the quantization axis --- this is the \textit{maximum} (or \textit{complete}) \textit{alignment} limit. As $\sigma/J$ increases, population spreads symmetrically across the magnetic substates toward the $|m| = J$ extremes, and in the limit of large $\sigma/J$ (i.e., as $\sigma/J \rightarrow \infty$) the distribution approaches the uniform statistical (isotropic) limit. \textit{Partial alignment} refers to the intermediate regime in which the symmetry condition $P(m) = P(-m)$ is satisfied but population is attenuated and distributed across multiple substates with $\sigma/J$ neither vanishingly small nor large.  

The general form of the $\gamma$-ray \textit{angular distribution} from an aligned nuclear state, measured as a function of the angle $\theta$ between the emitted $\gamma$-ray direction and the beam axis using a single detector, without any coincidence requirement, is given by
\begin{equation}
  W(\theta) = \sum_{k=0,2,4,\ldots} Q_{k}\, G_{k}(\tau)\, U_{k}\, B_{k}\, F_{k}\, P_{k}(\cos\theta),
  \label{eq:w-full}
\end{equation}
where the sum runs over even values of $k$ only. This expression should be distinguished from the $\gamma$-$\gamma$ \textit{angular correlation} function, in which $\theta$ denotes the angle between two successively detected $\gamma$ rays in a coincidence measurement and the alignment of the intermediate state is established by detection of the first radiation \cite{biedenharn:53, hamilton:40}; both take the same mathematical form, but the physical content of each factor differs. In Eq.~(\ref{eq:w-full}), the $F_k$ coefficients encode the nuclear structure information and are computed from appropriate Clebsch-Gordan and Racah coefficients following the formalism of Refs.~\cite{biedenharn:53, rose:67, yamazaki:67}, as described in Sect.~\ref{sec:method}. The $B_{k}$ factors are the statistical population tensors characterizing the degree of alignment of the initial state; $B_{0} = 1$ for all $J$ (see Sect.~\ref{sec:full_align}) and $B_{k} = 0$ for all $k > 0$ in the unoriented statistical limit, i.e., for isotropic distributions. The $Q_{k}$ factors are geometrical attenuation coefficients correcting for the finite solid angle subtended by the detector \cite{krane:73, krane:72}; a point detector would have $Q_{k} = 1$ for all $k$, while for a physical detector of finite size $Q_{k} < 1$ for $k>0$, with the attenuation becoming more pronounced for higher-ranked terms \cite{krane:72}.  Accurate determination of $Q_{k}$ may be non-trivial depending on the experimental configuration.  However, analytical approximations for various common source-detector geometries have been treated in a companion \texttt{Jupyter Notebook} and associated GitHub repository \cite{SolidAngles:GitHub}.  The $G_{k}(\tau)$ factors account for extranuclear perturbations based on hyperfine magnetic fields and electric quadrupole interactions with lattice field gradients that can disturb the nuclear spin orientation over the lifetime ($\tau$) of the state, thereby attenuating the measured anisotropy \cite{biedenharn:53, frauenfelder:53}.  For short-lived states, however, these effects are negligible and $G_{k} \approx 1$ \cite{smith:19}.  The $U_{k}$ factors are deorientation coefficients that account for the depopulation of the aligned state through unobserved intermediate transitions in a cascade as discussed in Sect.~\ref{sec:full_align}.  Finally, the $P_{k}(\cos\,\theta)$ terms are the Legendre polynomial functions of rank $k$.  For oriented states only even-rank statistical tensors are non-zero, and the $\gamma$-ray angular distribution involves only even-order Legendre polynomials.  This result follows from two independent constraints: the alignment condition itself eliminates odd-rank statistical tensors, and parity conservation in electromagnetic transitions independently suppresses odd-$k$ terms \cite{biedenharn:53}. It is also important to note that while $\gamma$-ray angular distributions are sensitive to the multipolarity order $L$ and the spin sequence $J_i \to J_f$, they cannot distinguish between electric (E$L$) and magnetic (M$L$) radiation of the same multipole order \cite{biedenharn:53, frauenfelder:53}; determination of the electromagnetic character requires additional measurements such as internal conversion coefficients or $\gamma$-ray linear polarization, e.g., Compton polarimetry \cite{morse:22}.  

In practice, following Yamazaki \cite{yamazaki:67}, the product $B_{k} F_{k}$ is frequently evaluated together and absorbed into a single anisotropy coefficient $A_{k} \equiv B_{k} F_{k}$, and Eq.~(\ref{eq:w-full}) reduces in the idealized limit, assuming negligible deorientation for a short-lived state measured with a point detector such that $Q_{k} = G_{k} = U_{k} = 1$, to the compact form 
\begin{equation}
  W(\theta) = \sum_{k=0,2,4,\ldots} A_{k}\, P_{k}(\cos\theta),
  \label{eq:w-reduced}
\end{equation}
where the $A_k$ anisotropy coefficients implicitly carry both the structural and alignment information of the initial state.

The theoretical framework underlying $\gamma$-ray angular distributions from aligned nuclei has a long history. The general theory of angular correlations of nuclear radiations was established by Hamilton \cite{hamilton:40} and developed into its modern form by Biedenharn and Rose \cite{biedenharn:53} using the statistical tensor formalism of Fano \cite{fano:57} and the Racah coefficient machinery \cite{racah:42}. The extension to angular distributions from reaction-oriented states was developed by Litherland and Ferguson \cite{litherland:61} and placed into a practically-convenient form by Krane, Steffen and Wheeler \cite{krane:73}, who introduced a \textit{directional-correlation-from-oriented-states} (DCO) functionality in terms of generalized $F$ coefficients. The foundational experimental methodology was reviewed by Frauenfelder \cite{frauenfelder:53}, while contemporary experiments continue to exploit the DCO ratio method, e.g., Ref.~\cite{ali:25}, enabling multipolarity characterization of emitted $\gamma$ rays and $J^{\pi}$ assignments of the levels involved.  The DCO method is a type of angular correlation analysis based on $\gamma$-$\gamma$ coincidence measurements where ratios of coincidence $\gamma$-ray intensities are measured at two different detector angles.  In experimental applications involving arrays of detectors with large solid-angle coverage, such as Gammasphere, DCO ratios are usually deduced for two different subsets of angles: detectors arranged at forward/backward polar angles with respect to the beam direction ("$\theta_{F/B} \approx 180^{\circ}$"), and those perpendicular to the beam ("$\theta_{\perp} \approx 90^{\circ}$").  Here, the DCO ratio (see, e.g., Refs.~\cite{deacon:11, steppenbeck:12}) is deduced as
\begin{equation}
    \label{eq:dco}
    R_{\text{DCO}} = \frac{I_{\gamma_{1}}^{(\theta_{F/B})} [\text{Gate}_{\gamma_{2}}^{(\theta_{\perp})}]}{I_{\gamma_{1}}^{(\theta_{\perp})} [\text{Gate}_{\gamma_{2}}^{(\theta_{F/B})}]},
\end{equation}
where $I_{\gamma_{1}}^{(\theta_{F/B})} [\text{Gate}_{\gamma_{2}}^{(\theta_{\perp})}]$ is the intensity of transition $\gamma_{1}$ in detectors at forward/backward angles $\theta_{F/B}$ when a coincidence-energy gate is set on transition $\gamma_{2}$ in detectors at perpendicular angles $\theta_{\perp}$, and similarly defined for $I_{\gamma_{1}}^{(\theta_{\perp})} [\text{Gate}_{\gamma_{2}}^{(\theta_{F/B})}]$.  A closely related variant, sometimes referred to as an \textit{angular-distribution-from-oriented-states} (ADO) ratio~\cite{piiparinen:96}, measures the ratio of single-transition intensities at two different angles (or two different subsets of angles) without requiring a coincidence gate on a second $\gamma$ ray:
\begin{equation}
    \label{eq:ado}
    R_{\text{ang}} = \frac{\Sigma_{\theta_{F/B}} I_{\gamma}}{\Sigma_{\theta_{\perp}} I_{\gamma}}.
\end{equation}
This approach is frequently adopted in cases where transition intensities or counting statistics are low.  In either case, the intensity ratios are calibrated against transitions of known multipolarity in well-studied nuclei providing a signature of the transition multipole order according to firmly-established limits for definitive $\Delta J=1$ (dipole) or $\Delta J=2$ (quadrupole) assignments.  However, as noted by Piiparinen \textit{et al}.~\cite{piiparinen:96}, what is often loosely, though sometimes misleadingly, referred to as a DCO ratio in the literature is more accurately understood as an angular distribution from the reaction-oriented nuclear state.  In this case, the alignment and angular anisotropy of $\gamma$-ray emission from the oriented state is exploited directly from the reaction dynamics rather than through a coincidence gate on a known transition.  In practice, the two methods are implemented similarly using asymmetric $\gamma$-$\gamma$ matrices in which one axis accepts events from all detectors regardless of angle and the other axis is incremented with events associated with a specific angular range, i.e., $\theta_{F/B} \approx 180^{\circ}$ or $\theta_{\perp} \approx 90^{\circ}$, and the distinction between them reduces to whether or not the all-angle axis is gated on a transition of known multipolarity.  Angular-intensity ratios deduced in this manner have become routine with large-detector arrays in recent years, e.g., the TESSA spectrometer at the Daresbury Laboratory \cite{rudolph:91}, the $8\pi$ array at Chalk River \cite{cromaz:99, chiara:01}, the HERA spectrometer at the Lawrence Berkeley National Laboratory (LBNL) \cite{bernstein:95}, the EXOGAM array at the Grand Acc{\'e}l{\'e}rateur National d'Ions Lourdes \cite{zheng:13}, the JUROGAM-II spectrometer at the University of Jyv{\"a}skyl{\"a} \cite{lv:21}, the Euroball-IV array at the Institut de Recherches Subatomiques \cite{buforn:00, jenson:02}, the Gammasphere array at LBNL \cite{lafosse:00, paul:07:I:II}, and the Gammasphere array at the Argonne National Laboratory \cite{clark:00, smith:01, zhu:06}.  However, not all of these references adhere strictly to the coincidence-gated DCO definition of Eq.~(\ref{eq:dco}).

Heavy-ion fusion-evaporation reactions provide particularly favorable conditions for the production of strongly aligned residual nuclei \cite{diamond:66, newton:67}. Projectile ions carry large orbital angular momenta into the compound system, leaving the angular momentum vector of the residual nucleus strongly aligned perpendicular to the beam axis following evaporation of a few neutrons. The neutron evaporation process generates an approximately Gaussian distribution of magnetic substates \cite{newton:67}, and the resulting pronounced $\gamma$-ray angular distribution anisotropies make these measurements a powerful spectroscopic tool for determining transition multipolarities and nuclear level spins and parities. This near-maximal alignment regime continues to be exploited in contemporary experimental programs, e.g., as demonstrated in low-energy inelastic proton scattering experiments using GRETINA \cite{morse:22} where the strongly aligned nuclear states produced enable determination of transition multipolarities and electromagnetic character.

For reactions involving lighter projectiles at lower incident energies, such as inelastic neutron-scattering measurements \cite{hurst:21, miriot:25}, the degree of alignment achieved is substantially less and a partial-alignment treatment is required. In such cases the Gaussian parametrization provides a practical and experimentally justifiable framework \cite{yamazaki:67, mateosian:74}, and $\gamma$-ray angular distribution measurements continue to provide important spectroscopic information \cite{lauritsen:25, longfellow:26, ali:25}. The same statistical tensor formalism has recently been applied in the context of nuclear fission, where Monte Carlo calculations have been used to quantify the spin alignment of fission fragments following neutron emission \cite{chalil:24}.

The resources most widely used in the analysis of experimental data have long been the \textit{Tables of Coefficients for Angular Distribution of Gamma Rays from Aligned Nuclei} by Yamazaki \cite{yamazaki:67}, the \textit{Angular Distributions of Gamma Rays in Terms of Phase-Defined Reduced Matrix Elements} by Rose and Brink \cite{rose:67}, and the companion partial-alignment tables of Der Mateosian and Sunyar \cite{mateosian:74}.  There are, however, significant practical limitations in the use of these tabulated resources.  The tables of Yamazaki \cite{yamazaki:67} provide the population tensor for ranks $k = 2,\,4,$ and 6, and the angular distribution coefficients for tensor ranks $k = 2$ and 4 only, covering multipole orders $1 \leq L \leq 3$ associated with initial integral spins $1 \leq J \leq 15\,\hbar$ and half-integral spins $3/2 \leq J \leq 31/2\,\hbar$.  Rose and Brink \cite{rose:67} extend the multipole coverage through to $L=4$ and the tensor rank to $k=2,\,4,\,6,$ and 8, but impose their own spin restrictions depending on the quantity in question: the angular distribution coefficients are tabulated up to highest initial spins of $J = 10\, \hbar$ (integral) and $J = 19/2\, \hbar$ (half-integral); the deorientation coefficients are given for multipole orders up to $L = 5$; the statistical tensor is tabulated for a maximum of seven magnetic substates, reaching spins of $J = 12 \,\hbar$ (integral) and $J = 23/2 \,\hbar$ (half-integral); and the population tensor is restricted to a maximum orbital angular momentum difference of $\Delta l=4$, with reaction-channel spin $0\leq s \leq 4$ and target spin $0 \leq J \leq 5\, \hbar$ for integral spins, or channel spin  $1/2 \leq s \leq 7/2$ and target spin up to $1/2 \leq J \leq 9/2\, \hbar$ for half-integral spins.  The partial-alignment tables of Der Mateosian and Sunyar \cite{mateosian:74} tabulate partial-alignment attenuation-anisotropy coefficients for tensor ranks $k=2$ and $k=4$ only, evaluated at a discrete set of Gaussian width parameters within the range $0.1 \leq \sigma/J \leq 2.0$, spanning integral spins over $1 \leq J \leq 26\,\hbar$ and half-integral spins from $3/2 \leq J \leq 51/2\,\hbar$, and restricted to dipole and quadrupole transitions only.  Fusion-evaporation reactions at modern stable ion-beam facilities routinely populate rotational cascades well beyond these ranges, rendering the tables inadequate in many contemporary cases. Moreover, these resources exist only in printed form, making their use inconvenient, potentially error prone, and no accompanying codebase was released alongside.

Several programs exist that address subsets of the functionality provided by the current project \texttt{PyGammaRAD}, including a lineage of increasingly capable Fortran implementations for the underlying vector-coupling coefficients \cite{tamura:70, wills:71, bretz:76, srinivasa:78, stone:80}, computer algebra system approaches based on proprietary software \cite{lai:94, Mathematica}, as well as more recent Java \cite{stevenson:02, COM:GitHub} and symbolic Python-based \cite{SymPy} open tools for the calculation of the coupling coefficients alone.  Macias \textit{et al}. \cite{macias:76} developed a Fortran program specifically for calculating $\gamma$-$\gamma$ directional correlation coefficients and mixing ratios.  This work was later modified by Rouse \textit{et al.} \cite{rouse:78} to include analysis of angular distributions for $\gamma$ rays of mixed multipolarities from partially-aligned nuclei.  However, these codes \cite{macias:76, rouse:78} are written in a language that is no longer straightforwardly accessible to many users and can only be used to calculate a limited subset of the angular distribution coefficients presented in the adopted literature \cite{yamazaki:67, mateosian:74, rose:67}.  Furthermore, none of the aforementioned capabilities integrates the vector-coupling machinery with the full apparatus of $\gamma$-ray angular distributions described in the foundational tabular works of Yamazaki \cite{yamazaki:67}, Rose and Brink \cite{rose:67}, and Der Mateosian and Sunyar \cite{mateosian:74}, in a single unified openly available package. \texttt{PyGammaRAD} addresses this gap: an open-source toolkit written in Python, it performs standard double-precision floating-point arithmetic and extended-precision arithmetic, as needed, that consolidates Clebsch-Gordan, Racah, Wigner $3$-$j$, $6$-$j$, $9$-$j$, and Gaunt coefficient calculations, together with the complete set of maximum \cite{yamazaki:67, rose:67} and partial \cite{yamazaki:67, mateosian:74} alignment angular distribution formalism presented in the associated canonical reference material.  To that end, \texttt{PyGammaRAD} is a self-contained tool based on conventional numerical computing that is portable, extensible, and accessible to the broad community of users familiar with the Python programming language and its associated libraries, making it straightforward to integrate into larger codebases. The program reproduces all tabulated data in Refs.~\cite{yamazaki:67, mateosian:74, rose:67} with full numerical precision --- in doing so, a small number of typographical errors in Ref.~\cite{rose:67} were identified --- and can readily evaluate angular-distribution coefficients for nuclei populated to arbitrarily high spin over a much wider range of multipolarities and tensor ranks than any of the original works.

A brief overview of the program structure is described in Sect.~\ref{sec:structure}. The formalism adopted for the calculation of $\gamma$-ray angular distribution coefficients is described in Sect.~\ref{sec:method}. Section~\ref{sec:check} provides a discussion of the benchmarking and validation procedures. In Sect.~\ref{sec:summary} a summary and outlook is presented. Finally, an overview of the underlying vector-coupling formalism commonly used in quantum physics applications based on the theory of angular momentum is presented in the \ref{sec:angmom}.

\section{\label{sec:structure}Structure of the program}


A Python implementation of a suite of calculators has been written to return precise analytic results for the angular distribution coefficients and tensors defined in Sect.~\ref{sec:method}, together with the associated angular momentum coupling coefficients which are required in the determination of these quantities.  The closely related angular momentum coupling symbols are also defined and coded into the suite, thus the complete software package offers broader application beyond the scope of the angular distribution focus of this manuscript.  The program is structured in a modular fashion and a single source file \texttt{am\_formulae.py} contains separate classes based on multiple inheritance for the angular momentum coupling coefficients: \texttt{am\_formulae.ClebschGordan}, \texttt{am\_formulae.Wigner3j}, \texttt{am\_formulae.Racah}, and \texttt{am\_formulae.Wigner9j} for the Clebsch-Gordan coefficient, Wigner-$3j$ symbol, Racah coefficient, and Wigner-$9j$ symbol calculators, respectively.  The Wigner-$6j$ symbol is strongly related to the Racah coefficient through a simple phase factor (see \ref{sec:6j}) and, therefore, a member function for its evaluation is also contained within the same \texttt{am\_formulae.Racah} class.  The calculation of the Wigner-$9j$ symbol is also simplified using \texttt{am\_formulae.Racah} class-inherited methods, as described in \ref{sec:9j}.  Gaunt coefficients can be decomposed into a product of two Wigner-$3j$ symbols (see \ref{sec:gaunt}) and can, therefore, be evaluated within the \texttt{am\_formulae.Wigner3j} class.  The callable methods used to return the exact solution of these vector coupling coefficients are then contained within \texttt{AngularMomentumCalculations} class of the \texttt{am\_methods.py} module.

The source file \texttt{angular\_distributions.py} contains the appropriate member functions of the \texttt{AngularDistributions} class needed to calculate the $\gamma$-ray angular distribution coefficients and tensors listed in the tabulated data of Yamazaki \cite{yamazaki:67}, as well as those in the tables of the appendix in the review article by Rose and Brink \cite{rose:67}, and the attenuated coefficients tabulated by Der Mateosian and Sunyar \cite{mateosian:74}.  To aid in the determination of the overall angular distribution calculation, the \texttt{angular\_distributions.Legendre} class contains the first 11 polynomials of the Legendre series coded in terms of $\cos \theta$.  Finally, the \texttt{tables.py} module contains classes and methods for handling and manipulating the tabulated data of Refs.~\cite{yamazaki:67, mateosian:74, rose:67}.  Effectively, this module serves as an API for interacting with the published tables in these works and also allows users to fully regenerate the tables in convenient machine-readable CSV and JSON formats.

All of these classes are consolidated through multiple inheritance into a single top-level \texttt{AngularMomentum} class, which serves as the unified user-facing API to the complete \texttt{PyGammaRAD} library.  Consequently, once an instance of \texttt{AngularMomentum} is created, all methods --- regardless of which source module they originate from --- are directly accessible through that single instance, as illustrated in the next Sect.~\ref{sec:running}

\subsection{\label{sec:running}Running the program}

The \texttt{PyGammaRAD} software package is open source, freely available on GitHub \cite{PyGammaRAD:GitHub} and the Python Package Index (PyPI) repository \cite{PyGammaRAD:PyPI}, and comes packaged with a permissive MIT license.  Building and installation are straightforward and are detailed in the accompanying \texttt{README} documentation on the GitHub and PyPI landing pages.  The calculation of all quantities and utilization of associated functions is enabled by creating an instance of the \texttt{AngularMomentum} class after importing the \texttt{PyGammaRAD} library, for example:
\begin{lstlisting}[language=Python, morekeywords={as, In, Out}]
  In [1]: import PyGammaRAD as pg
  In [2]: am = pg.AngularMomentum()
  
  # Calculate Clebsch-Gordan coefficient <j1=5/2 m1=3/2 j2=5/2 m2=-1/2 | j=1 m=1>
  In [3]: am.cg(2.5, 1.5, 2.5, -0.5, 1, 1)
  Out[3]: -0.4780914437337574 # -2*sqrt(2/35)
  
  # Calculate coefficient F(k=2,Jf=0,L1=2,L2=2,Ji=2)
  In [4]: am.calc_F(2, 0, 2, 2, 2)
  Out[4]: -0.5976143046671966

  # Compare to result in "Table 2(a)" [Yamazaki 1967]
  In [5]: am.get_row_table2(2,0,2,2,2,coeff='F')
  Out[5]: -0.59761
  
  # Print "Table 2(a)" from Ref. [Yamazaki 1967] to file in JSON format
  In [6]: am.yamazaki2file('T2A', 'JSON')
  INFO:PyGammaRAD.log_handlers:YamazakiTable2a.json
\end{lstlisting}
To help illustrate various workflows and the overall utility of the software, including verification of methods against published results (see Sect.~\ref{sec:check}), \texttt{PyGammaRAD} comes shipped with several different \texttt{Jupyter Notebooks} available from the project GitHub repository \cite{PyGammaRAD:GitHub}.  All \texttt{PyGammaRAD} classes and member functions have supporting docstrings that can be invoked at the console as described in the \texttt{README}.  These docstrings provide a short explanation of the function, any arguments passed, the return value, as well as exceptions that are raised by the function.  Each docstring is also presented with at least one known working example to showcase the corresponding method execution.  All angular distribution and angular momentum coupling methods available to the \texttt{PyGammaRAD} library are also summarized in the associated \texttt{README} documentation, along with methods enabling retrieval and manipulation of the tabulated data in Refs.~\cite{yamazaki:67, mateosian:74, rose:67}.

\subsection{\label{sec:logtest}Logging, testing, and dependencies}

A logging framework has been implemented within \texttt{PyGammaRAD} whereby messages are categorized appropriately using different levels of severity.  By default, logs are only directed to the console; however, an application log can also be redirected to file upon request.  The implemented logging system provides a centralized and standardized way to handle the field reporting of application events such as errors and warnings.  The logging system is configured to provide useful information and metadata regarding such events, including timestamps, module names, function names, and line numbers, thus enabling a richer context for debugging and analysis, making it easier for the user to monitor and troubleshoot issues.

A suite of Python modules comprising more than 200 unit tests has been developed as part of this project to help validate and benchmark its performance against published results, verification of return types, and raising of exceptions when certain conditions arise.  All angular distribution coefficients and tensors presented in Sect.~\ref{sec:method}, together with the angular momentum coupling coefficients presented in ~\ref{sec:angmom}, have dedicated modules that comprise the test suite.  The test scripts are run to ensure compliance with the local Python environment and have been successfully tested across multiple Python versions 3.6 to 3.13 inclusive on both Linux and Mac OS platforms.  The \texttt{PyGammaRAD} software package has the following third-party Python-package dependencies: \texttt{numpy}, \texttt{pandas}, and \texttt{pytest}.  There is also an additional \texttt{scipy} package dependency to run certain actions in the accompanying notebooks.

\section{\label{sec:method}Formalism used in the calculations}

The calculators used in the evaluation of the overall angular distribution coefficients and tensors are based on the well-known sum formulae for the coupling of angular momenta vectors which have been extensively described in various dedicated texts on the subject matter, e.g., Refs.~\cite{brink:68, devanathan:02, edmonds:57, rose:57}, as well as in the published literature, e.g., Refs.~\cite{tamura:70, wills:71, bretz:76, srinivasa:78, stone:80, lai:94, wei:98, rasch:03, stevenson:02, johansson:16}.  The vector-coupling routines are described in \ref{sec:angmom}, while the quantities explicitly of relevance to angular distributions are described in the following subsections.

\subsection{\label{sec:angdist}$\gamma$-ray angular-distribution coefficient calculator}

Methods have been developed based on the formalism outlined by Yamazaki \cite{yamazaki:67} and Rose and Brink \cite{rose:67} to calculate angular distribution functions in maximally-aligned nuclei.  Additionally, we have further developed methods for partial alignment in nuclei based on the formalism outlined by Der Mateosian and Sunyar \cite{mateosian:74}, which in itself is an extension of the earlier work by Yamazaki \cite{yamazaki:67}.  Here, we present an overview of the corresponding expressions used to calculate the various functions and coefficients that are needed to reproduce all the tabulated data in Refs.~\cite{yamazaki:67, mateosian:74, rose:67}.  Furthermore, these methods can be applied for values of $J$, $L$, $m$, and $k$ well beyond the ranges presented in the original works of Yamazaki \cite{yamazaki:67}, Rose and Brink \cite{rose:67}, and Der Mateosian and Sunyar \cite{mateosian:74}.  

Two important special cases are worth noting.  Firstly, for states characterized by $J=0$ the single magnetic substate $m=0$ means no preferred direction can be established and emission is necessarily isotropic.  Secondly, and for states with $J=1/2 \, \hbar$, the two magnetic substates $m= \pm 1/2$ are equally populated under reaction-symmetry conditions imposed by the reflection symmetry about the beam axis, forcing a symmetric distribution through the $m = \pm 1/2$ substates due to the orbital angular momentum of the incoming projectile, which again gives rise to an isotropic distribution.  Accordingly, in the canonical reference material \cite{yamazaki:67, mateosian:74, rose:67}, there are no initial states with $J_{i} = 0$ or $J_{i} = 1/2 \, \hbar$ since the associated anisotropy coefficients are expected to vanish in these cases.

The adopted formalism based on complete and partial alignment are discussed, in turn, below.

\subsubsection{\label{sec:full_align} Maximum nuclear alignment}

The overall angular distribution function \cite{yamazaki:67} for a $\gamma$-ray transition from an initial state to a final state $J_{i} \rightarrow J_{f}$ may be adequately expressed using the leading-order terms from Eq.~(\ref{eq:w-reduced}), thus
\begin{equation}
  \label{eq:w-1}
  W(\theta) = 1 + A_{2}P_{2}(\cos \theta) + A_{4}P_{4}(\cos \theta),
\end{equation}
where $A_{k}$ are the anisotropy coefficients and $P_{k}$ are the Legendre polynomials of order $k$.  In principle, the series extends to higher $k$ as per Eqs.~(\ref{eq:w-full}) and (\ref{eq:w-reduced}) earlier, however, the truncation at $k=4$ is conventional for the cases addressed by the original tables \cite{yamazaki:67} since higher-rank terms vanish for low spins or are negligible.  We should also note that \texttt{PyGammaRAD} is not limited in this way and distributions can be straightforwardly calculated for higher-rank $k>4$ terms.

To obtain $A_{k}$ in Eq.~(\ref{eq:w-1}), we first need to calculate the statistical population tensor $B_{k}$ and the $F_{k}$ angular distribution coefficient \cite{yamazaki:67}.  In the case of complete nuclear alignment, we can write the statistical population tensor for integral and half-integral spin, respectively, as
\begin{align}
  \label{eq:w-2}
  \notag
  B_{k}(J) &= (-1)^{J} \sqrt{2J+1} \langle J 0 J 0 | k 0 \rangle; \\
  B_{k}(J) &= (-1)^{J-\frac{1}{2}} \sqrt{2J+1} \langle J \frac{1}{2} J -\frac{1}{2} | k 0 \rangle,
\end{align}
where $\langle J 0 J 0 | k 0 \rangle$ and $\langle J \frac{1}{2} J -\frac{1}{2} | k 0 \rangle$ are Clebsch-Gordan coefficients for given $J$ and $k$.  The $F_{k}$ angular distribution coefficient is given as
\begin{align}
  \label{eq:w-3}
  \notag
  F_{k}(J_{f}L_{1}L_{2}J_{i}) &= (-1)^{J_{f}-J_{i}-1}\sqrt{(2L_{1}+1)(2L_{2}+1)(2J_{i}+1)} \\
  &\times \langle L_{1} 1 L_{2} -1 | k 0 \rangle W(J_{i}J_{i}L_{1}L_{2}; kJ_{f}),
\end{align}
where $W(J_{i}J_{i}L_{1}L_{2}; kJ_{f})$ is a Racah coefficient, and $L_{1}$ and $L_{2}$ are the angular momentum multipole orders for a mixed transition whereupon the lowest two orders are considered such that the interfering multipole $L_{2} = L_{1} + 1$.  In the case of a pure transition $L_{1} = L_{2} = L$.  For the ideal case of maximal alignment we can then compute $A_{k}$ using
\begin{align}
  \label{eq:w-4}
  \notag
  A_{k}^{\text{max}}(J_{i}L_{1}L_{2}J_{f}) &= \frac{1}{1+\delta_{\gamma}^{2}} \left[B_{k}(J_{i})F_{k}(J_{f}L_{1}L_{1}J_{i}) + 2\delta_{\gamma} B_{k}(J_{i})F_{k}(J_{f}L_{1}L_{2}J_{i}) \right. \\
    &+ \left. \delta_{\gamma}^{2}B_{k}(J_{i})F_{k}(J_{f}L_{2}L_{2}J_{i}) \right],
\end{align}
where $\delta_{\gamma}$ is the $\gamma$-ray mixing ratio \cite{krane:75}.  The mixing ratio can defined in terms of the reduced matrix elements of the associated electromagnetic multipole operators:
\begin{equation}
    \label{eq:mixing_ratio}
    \delta_{\gamma} = \frac{\langle J_{f} || \mathcal{M}(L_{2}) || J_{i} \rangle}{\langle J_{f} || \mathcal{M}(L_{1}) || J_{i} \rangle}.
\end{equation}
In the case of a pure transition when $L_{1} = L_{2}$ there is no mixing and $\delta_{\gamma} = 0$.  For other cases where $|\delta_{\gamma}| > 0$, $\delta_{\gamma}$ should be obtained from experiment, e.g., \cite{ali:25, krane:70}, and provided as an input parameter to \texttt{PyGammaRAD} methods as appropriate.  A survey of $M1+E2$ mixing ratios from experimental angular distribution and correlation data is described and the results are tabulated in Ref.~\cite{krane:75}. Alder and Steffen \cite{alder:75} provide a comprehensive discussion of electromagnetic multipole operators and their associated matrix-element transformation properties.

The anisotropy coefficient, expressed in Eq.~(\ref{eq:w-3}) above, can alternatively be described in terms of the closely related $R_{k}$ angular distribution coefficient \cite{rose:67}:
\begin{align}
  \label{eq:R1}
  \notag
  R_{k}(L_{1}L_{2}J_{i}J_{f}) &= (-1)^{1+J_{i}-J_{f}+L_{2}-L_{1}-k}\sqrt{(2L_{1}+1)(2L_{2}+1)(2J_{i}+1)} \\
  &\times \langle L_{1} 1 L_{2} -1 | k 0 \rangle W(J_{i}J_{i}L_{1}L_{2}; kJ_{f}).
\end{align}
Clearly, Eqs.~(\ref{eq:w-3}) for $F_{k}$ and (\ref{eq:R1}) for $R_{k}$ differ only in their phases and can be related through the following expression
\begin{equation}
  \label{eq:R2}
  R_{k}(L_{1}L_{2}J_{i}J_{f}) = (-1)^{L_{1}-L_{2}+k}F_{k}(J_{f}L_{1}L_{2}J_{i}),
\end{equation}
and so the anisotropy coefficient given in Eq.~(\ref{eq:w-4}) may then be recast by replacing the $F_{k}$ coefficients with their corresponding $R_{k}$ coefficients \cite{rose:67}, thus allowing for the overall distribution function [Eq.~(\ref{eq:w-1})] to be expressed in terms of $R_{k}$ instead.

In a $\gamma$-ray cascade, for a state with spin $J_{f}$ whose formation is dependent entirely upon the preceding $J_{i} \rightarrow J_{f}$ transition, it is useful to introduce the $U_{k}$ angular distribution coefficient \cite{yamazaki:67}, often referred to as the deorientation (or depolarization) coefficient \cite{krane:71}, which describes the effect of unobserved transitions preceding the observed radiation.  This coefficient has the effect to alter the magnetic substate populations thereby reducing the degree of orientation ($U_{k} < 1$).  It is dependent upon the multipolarities of the unobserved transitions, together with the angular momenta of the intermediary states, and may be written as
\begin{equation}
  \label{eq:w-8}
  U_{k}(J_{i}L_{1}L_{2}J_{f}) = \frac{1}{1+\delta_{\gamma}^{2}}\left[u_{k}(J_{i}L_{1}J_{f}) + \delta_{\gamma}^{2}u_{k}(J_{i}L_{2}J_{f})\right],
\end{equation}
where
\begin{equation}
  \label{eq:w-9}
  u_{k}(J_{i}L_{1}J_{f}) = (-1)^{J_{i}+J_{f}-L_{1}} \sqrt{(2J_{i}+1)(2J_{f}+1)} W(J_{i}J_{i}J_{f}J_{f}; kL_{1}).
\end{equation}
Alternatively, Eq.~(\ref{eq:w-9}) above may also be expressed as a ratio of Racah coefficients \cite{rose:67} given by
\begin{align}
    \label{eq:racah-ratio}
    \notag
    u_{k}(J_{i}L_{1}J_{f}) &= (-1)^{k} \frac{W(J_{i}J_{f}J_{i}J_{f}; L_{1}k)}{W(J_{i}J_{f}J_{i}J_{f}; L_{1}0)} \\
    &= (-1)^{k}\frac{W(J_{i}J_{i}J_{f}J_{f}; kL_{1})}{W(J_{i}J_{i}J_{f}J_{f}; 0L_{1})}.
\end{align}

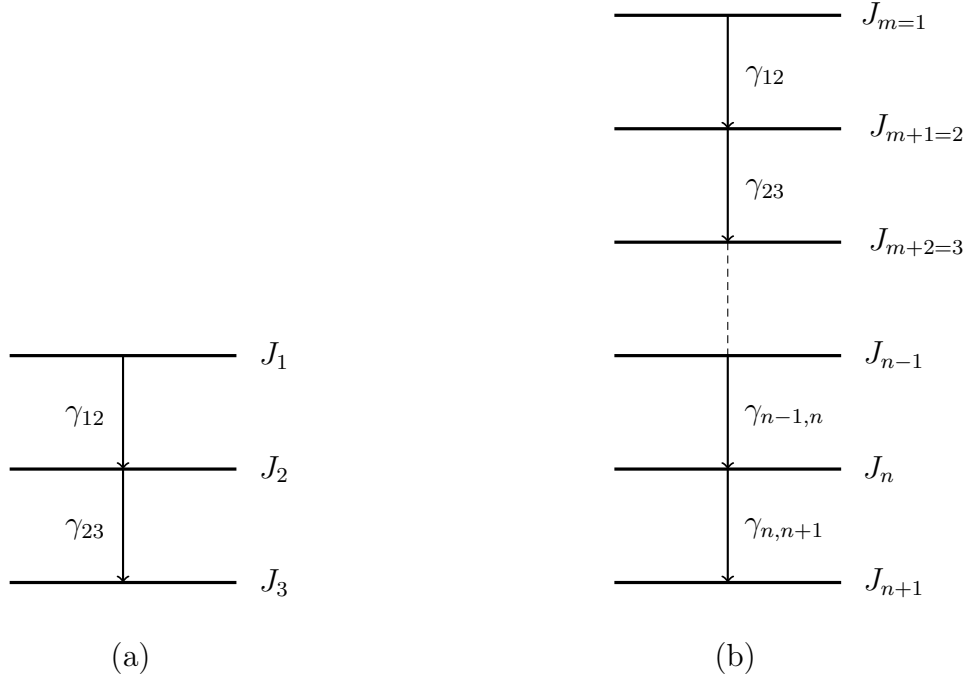
\begin{figure}[t!]
    \centering 
    \begin{tikzpicture}
        \draw[very thick] (0, 5.0) -- (3, 5.0);
        \draw[very thick] (0, 3.5) -- (3, 3.5);
        \draw[very thick] (0, 2.0) -- (3, 2.0);

        \draw[thick,->] (1.5, 5.0) -- (1.5, 3.5);
        \draw[thick,->] (1.5, 3.5) -- (1.5, 2.0);

        \draw (3.5, 5.0) node {$J_{1}$};
        \draw (3.5, 3.5) node {$J_{2}$};
        \draw (3.5, 2.0) node {$J_{3}$};

        \draw (1.0, 4.2) node {$\gamma_{12}$};
        \draw (1.0, 2.7) node {$\gamma_{23}$};

        \draw (1.6, 1.0) node {(a)};

        \draw[very thick] (8.0, 9.5) -- (11.0, 9.5);
        \draw[very thick] (8.0, 8.0) -- (11.0, 8.0);
        \draw[very thick] (8.0, 6.5) -- (11.0, 6.5);
        \draw[very thick] (8.0, 5.0) -- (11.0, 5.0);
        \draw[very thick] (8.0, 3.5) -- (11.0, 3.5);
        \draw[very thick] (8.0, 2.0) -- (11.0, 2.0);

        \draw[thick,->] (9.5, 9.5) -- (9.5, 8.0);
        \draw[thick,->] (9.5, 8.0) -- (9.5, 6.5);
        \draw[densely dashed] (9.5, 6.5) -- (9.5, 5.0);
        \draw[thick,->] (9.5, 5.0) -- (9.5, 3.5);
        \draw[thick,->] (9.5, 3.5) -- (9.5, 2.0);

        \draw (11.7, 9.5) node {$J_{m=1}$};
        \draw (12.0, 8.0) node {$J_{m+1=2}$};
        \draw (12.0, 6.5) node {$J_{m+2=3}$};
        \draw (11.7, 5.0) node {$J_{n-1}$};
        \draw (11.5, 3.5) node {$J_{n}$};
        \draw (11.7, 2.0) node {$J_{n+1}$};

        \draw (10.0, 8.7) node {$\gamma_{12}$};
        \draw (10.0, 7.2) node {$\gamma_{23}$};
        \draw (10.25, 4.2) node {$\gamma_{n-1, n}$};
        \draw (10.25, 2.7) node {$\gamma_{n, n+1}$};

        \draw (9.6, 1.0) node {(b)};
    \end{tikzpicture}
    \caption{\label{fig:cascade} (a) Two-step $\gamma$-ray cascade. (b) Multi-step $\gamma$-ray cascade.}
\end{figure}

If we consider the two-step $\gamma$-ray cascade involving transitions between states from $J_{1} \rightarrow J_{2} \rightarrow J_{3}$ as shown in Fig.~\ref{fig:cascade}a, where $J_{1}$ is the initially populated aligned state satisfying the alignment condition $P_{m_{i}}(J_{1}) = P_{-m_{i}}(J_{1})$, with the normalization condition $\sum\limits_{m_{i}=-J_{1}}^{+J_{1}}P_{m_{i}}(J_{1})=1$ satisfied, and characterized by statistical tensors $B_{k}(J_{1}) \neq 0$ for even $k>0$, the distribution formula expressed in Eq.~(\ref{eq:w-1}) will need to be modified for the second $\gamma$ ray, labeled $\gamma_{23}$ in Fig.~\ref{fig:cascade}a, which then has the following distribution \cite{rose:67}
\begin{equation}
  \label{eq:w-mod1}
  W(\theta) = \sum\limits_{k=0,2,4} U_{k}(J_{i}=J_{1},J_{f}=J_{2}) A_{k}(J_{i}=J_{2},J_{f}=J_{3})P_{k}(\cos \theta).
\end{equation}
Equation~(\ref{eq:w-mod1}) can be generalized for the $n^{\text{th}}$ $\gamma$ ray in a multi-step cascade, as depicted in Fig.~\ref{fig:cascade}b, by including a product of $U_{k}$ terms for each preceding non-observed $\gamma$ ray, i.e., all $\gamma$ rays other than $\gamma_{n,n+1}$ shown in Fig.~\ref{fig:cascade}b, such that
\begin{align}
  \label{eq:w-mod2}
  \notag
  W(\theta) &= \sum\limits_{k=0,2,4} \Big[ A_{k}(J_{i}=J_{n},J_{f}=J_{n+1})P_{k}(\cos \theta) \\
    & \times  \prod\limits_{J_{m},J_{m+1}}^{J_{n-1},J_{n}} U_{k}(J_{i}=J_{m},J_{f}=J_{m+1}) \Big].
\end{align}

Another important quantity that arises in the treatment of $\gamma$-ray angular distributions is the statistical tensor.  Rose and Brink \cite{rose:67} define this tensor in terms of both spin $J$ and its quantized magnetic substate projection $m$, thus
\begin{equation}
  \label{eq:p}
  \rho_{k}(J,m) = (2-\delta_{m,0}) \sqrt{2J+1} (-1)^{J-m} \langle J m J -m | k 0 \rangle,
\end{equation}
this equation, therefore, represents the tensor element for a definite magnetic substate $m$ without population weighting.  Here, the Kronecker delta function $\delta_{m,0} = 1$ for $m = 0$; otherwise $\delta_{m,0} = 0$ for all other values of $m$.  This expression removes the dependency of any knowledge of the associated population parameters \cite{yamazaki:67}, and accordingly, $\rho_{k}(J,m)$ can then be calculated for pure individual $m$-substate projections for a given $J$.

The final coefficient that we consider in this section is the $S_{k}$ coefficient \cite{rose:67} which characterizes the alignment of a nuclear state $J$ when that state is formed in a particle-induced reaction and is written as
\begin{align}
  \label{eq:S}
  \notag
  S_{k}(l_{1}l_{2}J s) &= (-1)^{s-J}\sqrt{(2l_{1}+1)(2l_{2}+1)(2J+1)} \\
  &\times \langle l_{1} 0 l_{2} 0 | k 0 \rangle W(J J l_{1}l_{2}; k s).
\end{align}
Here $l_{1}$ and $l_{2}$ are the orbital angular momenta of the incident particle relative to the target nucleus --- two interfering partial waves representing distinct ways the incident particle can couple its orbital motion to form the same compound-nucleus state $J$ --- and $s$ denotes the reaction-channel spin which corresponds to the vector sum of the spin of the incident particle and the spin of the target nucleus.  All three entrance-channel quantities, i.e., $l_{1}$, $l_{2}$, and $s$, therefore pertain to the incident particle and target nucleus, rather than the exit channel of the reaction.  The $S_k$ coefficient encodes the dependence of the alignment coefficients $B_{k}(J)$ on this partial-wave content (note that $S_{k}$ is an input to $B_{k}(J)$ \cite{rose:67}, not a replacement for it; both are referred to as population tensors in the literature but they are distinct quantities). This formalism applies to resonant capture reactions such as 
$(p,\gamma)$, $(n,\gamma)$, and $(\alpha,\gamma)$, as well as to inelastic-scattering reactions with significant compound-nucleus contributions, such as $(p,p')$ and $(n,n')$. In this respect $S_{k}$ is distinct in character from $R_{k}$ and $U_{k}$: rather than characterizing the observed $\gamma$-ray transition or an unobserved cascade step, it describes the reaction mechanism responsible for establishing the initial nuclear alignment.

\subsubsection{\label{sec:partial_align} Partial nuclear alignment}

\begin{figure}[t!]
    \centering
    \begin{subfigure}[b]{0.48\textwidth}
        \centering
        \includegraphics[width=\textwidth]{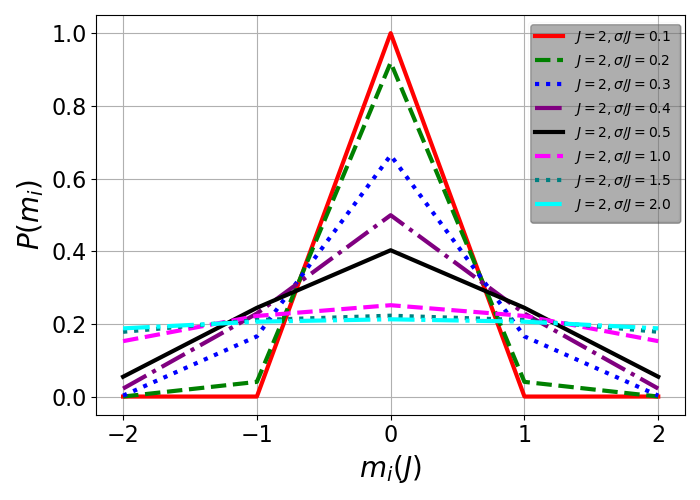}
        \caption{$J=2\,\hbar$}
        \label{fig:J2}
    \end{subfigure}
    \hfill
    \begin{subfigure}[b]{0.48\textwidth}
        \centering
        \includegraphics[width=\textwidth]{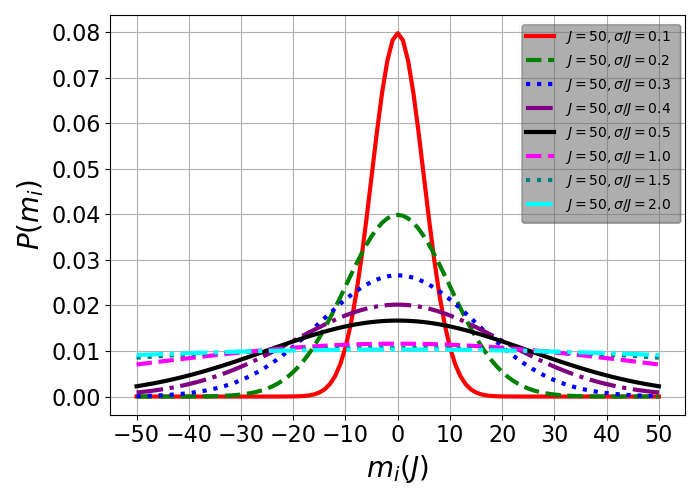}
        \caption{$J=50\,\hbar$}
        \label{fig:J50}
    \end{subfigure}
    \caption{\label{fig:mpops} Magnetic substate populations $P(m_{i})$ as a function of $m_{i}(J)$, shown for aligned (a) $J=2\,\hbar$ and (b) $J=50\,\hbar$ states.  In each case, the Gaussian-width control parameter is varied from $\sigma/J=0.1$, where the distributions approach maximal alignment with $P(m_{i})$ concentrated on $m=0$, to $\sigma/J=2.0$, where the distributions approach an unoriented uniform distribution for all $m$.}
\end{figure}

In real experimental scenarios, the degree of alignment is often only partial \cite{ali:25, longfellow:26, lauritsen:25} and modifications to the overall distribution function given by Eq.~(\ref{eq:w-1}) need to be accounted for.  Here, an attenuation coefficient is defined such that
\begin{equation}
  \label{eq:w-5}
  \alpha_{k}(J) = \frac{\rho_{k}(J)}{B_{k}(J)},
\end{equation}
where $\rho_{k}$ represents the statistical tensor \cite{yamazaki:67} for a state $J$ defined by the population parameters $P_{m}$ that specify the formation process for this state and is given by
\begin{equation}
  \label{eq:w-6}
  \rho_{k}(J) = \sqrt{2J+1} \sum\limits_{m=-J}^{J} (-1)^{J-m} \langle J m J -m | k 0 \rangle P_{m}(J).
\end{equation}
The expression embodied by Eq.~(\ref{eq:w-6}) represents the full-weighted statistical tensor and differs from the earlier Eq.~(\ref{eq:p}), which describes the statistical tensor as a function of $J$ and $m$ individually, in that this expression is summed over all values of $m$ weighted by their corresponding population parameter.  In cases where the alignment is only partial, the states are oriented with attenuated magnetic-substate populations although the symmetry about $m=0$ still holds, i.e., $P_{m}(J) = P_{-m}(J)$, and $\rho_{k}(J)$ vanishes for odd-$k$.  If $P_{m}(J)$ is unknown, then $\rho_{k}(J)$ given by Eq.~(\ref{eq:w-6}) in the work of Yamazaki \cite{yamazaki:67} is, in general, difficult to evaluate.  Experimentally, however, it is often justifiable to assume a Gaussian distribution of magnetic substates \cite{yamazaki:67, mateosian:74, lauritsen:25} characterized by the half-width parameter $\sigma$ such that the $m$-substate population parameter may be calculated in accordance with the following expression
\begin{equation}
    \label{eq:pp}
    P_{m}(J) = \frac{\exp\left(-\frac{m^{2}}{2\sigma^{2}}\right)}{\sum\limits_{m=-J}^{J}\exp\left(-\frac{m^{2}}{2\sigma^{2}}\right)}.
\end{equation}

Under this assumption, the dependence of the partial-alignment attenuation coefficient $\alpha_{k}(J)$ is controlled by the Gaussian-width distribution parameter $\sigma/J$, and the modified angular distribution function may then be written as
\begin{equation}
    \label{eq:w-partial}
    W(\theta) = 1 + \alpha_{2}A_{2}P_{2}(\cos \theta) + \alpha_{4}A_{4}P_{4}(\cos \theta).
\end{equation}
Again, the explicit sum over the $k = 0,\, 2,\, 4$ terms reflects the practical truncation imposed in the original tables \cite{mateosian:74}; however, \texttt{PyGammaRAD} is not restricted to $k\leq4$.  The $\alpha_{k}A_{k}$ term represents the attenuated anisotropy coefficient that can be compared with experimental anisotropy coefficients, frequently labeled $a_{k}$ \cite{ali:25, longfellow:26, lauritsen:25}.  Furthermore, in the analysis of such experimental data, the multipole mixing ratio $\delta_{\gamma}$ --- defined earlier by Eq.~(\ref{eq:mixing_ratio}) in Sect.~\ref{sec:full_align} --- is typically extracted by comparing the measured ratio $a_{2}/a_{4}$ of angular distribution coefficients with theoretically-calculated values for a given $\sigma/J$ and interpolating to find the value of $\delta_{\gamma}$ consistent with the experimental ratio \cite{mateosian:74, ali:25}.

For illustrative purposes, in Fig.~\ref{fig:mpops} we show the transition from near-complete alignment in $J=2 \,\hbar$ and $J=50 \,\hbar$ states centered on $m=0$ ($\sigma/J=0.1$), with vanishingly small populations in the corresponding $|m| > 0$ substates, to a near purely-statistical unoriented distribution revealing approximately uniform populations across all $m$ ($\sigma/J=2.0$).  The attenuation in alignment as $\sigma/J$ increases is also clear in this figure; various degrees of partial alignment in the intermediate region between these two limits are also shown.

\section{\label{sec:check}Verification of the programs}

The methods presented in Sect.~\ref{sec:angdist} were checked against the tabulated angular distribution coefficients in the data tables of Yamazaki \cite{yamazaki:67}, Rose and Brink \cite{rose:67}, and Der Mateosian and Sunyar \cite{mateosian:74}.  In all cases, other than a few typographical errors that were found in the $R_{k}$ and $\rho_{k}$ tables of Rose and Brink \cite{rose:67},\footnote{Typographical errors discovered in the angular distribution $R_{k}$ [Eq.~(\ref{eq:R1})] and statistical tensor $\rho_{k}$ [Eq.~(\ref{eq:p})] coefficients tabulated in the appendix of Rose and Brink \cite{rose:67} by comparison with \texttt{PyGammaRAD} \cite{PyGammaRAD:GitHub, PyGammaRAD:PyPI}.
\begin{center}
\begin{tabular}{lcc}
\hline\hline
Coefficient & Rose and Brink \cite{rose:67} & \texttt{PyGammaRAD} \cite{PyGammaRAD:GitHub, PyGammaRAD:PyPI}\\
\hline
$R_{k=6}(L_{1}=3\,L_{2}=4\,J_{i}=3\,J_{f}=6)$ & $-0.0663$ & $-0.0063$ \\
$R_{k=2}(L_{1}=2\,L_{2}=2\,J_{i}=17/2\,J_{f}=13/2)$ & 0.3776 & $-0.3776$ \\
$\rho_{k=2}(J=4\, m=0)$ & $-1.1366$ & $-1.1396$ \\
\hline\hline
\end{tabular}
\end{center}
} the published data are reproduced precisely, which further serves as an implicit verification of the vector-coupling calculation machinery underpinning the evaluation of the Clebsch-Gordan and Racah coefficients that are needed in the overall determination of the various calculated angular-distribution coefficients and statistical tensors.  As a further check, the Clebsch-Gordan coefficients and Wigner 3-$j$, 6-$j$, and 9-$j$ symbols, based on the formalism outlined in \ref{sec:angmom}, were checked against all tabulated values published by Stevenson~\cite{stevenson:02}.  Complete agreement with the tabulated values of Stevenson \cite{stevenson:02} was obtained with the exception of a single Wigner-$9j$ entry; the affected entry appears to be a typographical error which is detailed in the supplementary \texttt{Jupyter Notebook} \cite{PyGammaRAD:GitHub} and no correction to the underlying formalism \cite{stevenson:02, COM:GitHub} is implied.  We also checked the calculation of Wigner-$9j$ symbols with \texttt{Mathematica} \cite{Mathematica} as well as tabulated results published by Tamura \cite{tamura:70} and Wei \cite{wei:98}.  Clebsch-Gordan and Racah coefficients tabulated by Tamura \cite{tamura:70} were also confirmed using our calculator.  Additionally, we were able to confirm the calculation of the angular momentum couplings against the online Wigner coefficient calculator based on the root-rational-fraction program developed by Stone~\cite{stone:80}.  Further tests were also performed and confirmed against the \texttt{SymPy} symbolic mathematics library \cite{SymPy}, which provides a complete suite of vector-coupling functions, including Clebsch-Gordan, Racah, Wigner 3-$j$, 6-$j$, 9-$j$, and Gaunt coefficients, within a modern open Python environment.  The published literature on tabulated Gaunt coefficients is more limited; although we were generally able to reproduce the results reported by Y{\"u}kc{\"u} \cite{yukcu:25}.  However, a number of errors were found in the proposed couplings of this work due to a failure to satisfy the condition $\Sigma_{m_{i}=1}^{m_{i}=3} m_{i} = 0$.  These coupling errors are addressed in the accompanying \texttt{Jupyter Notebook} \cite{PyGammaRAD:GitHub} and we have confirmed our results by comparison with \texttt{SymPy} \cite{SymPy}.

The companion \texttt{Jupyter Notebooks} distributed with this project \cite{PyGammaRAD:GitHub} also contain actions confirming all the aforementioned published results \cite{yamazaki:67, mateosian:74, rose:67, stevenson:02, stone:80, tamura:70, wei:98, yukcu:25}, thus providing the user with convenient reference workflows and a means of inspection to perform independent checks of the data and calculation methods.  Furthermore, as part of the test-driven development process, an extensive suite of unit tests has also been written.  These unit tests were, in part, developed to corroborate the well-established results of Refs.~\cite{yamazaki:67, mateosian:74, rose:67, stevenson:02, stone:80, tamura:70, wei:98, yukcu:25}.  Unit tests have been written to establish the correct behavior for the complete apparatus of angular-distribution formalism presented in Sect.~\ref{sec:angdist} as well as the angular momentum formulae of \ref{sec:angmom} upon which vector-coupling suite of \texttt{PyGammaRAD} is based.  It is intended for users to run through the test script to demonstrate the established conformity of all results within their local environment.

We can also assert that Eq.~(\ref{eq:w-9}) \cite{yamazaki:67} and Eq.~(\ref{eq:racah-ratio}) \cite{rose:67} are equivalent formal methods for determining the $u_{k}$ angular distribution coefficient.  This claim was checked by performing an extensive set of unit tests over all integral and half-integral results presented in the data tables of Yamazaki \cite{yamazaki:67}.  These tests are also distributed as part of the test suite.

 Angular momentum couplings based on large integer arguments have also been tested.  \texttt{PyGammaRAD} reproduces all very large angular momenta couplings reported by Stevenson \cite{stevenson:02} as well as those by Johansson and Forss{\'e}n \cite{johansson:16} -- with the exception of the most extreme cases involving $j=10,000$ or $j=50,000$ \cite{johansson:16}, for which \texttt{PyGammaRAD} is not a suitable tool.  The built-in integer type of the Python programming language automatically handles arbitrarily large integers, so there is no inherent size limit beyond the available memory.  However, for certain very large integers, for example, factorials of large numbers, the native \texttt{math.sqrt} method may not execute the conversion to a floating-point representation, resulting in an \texttt{OverflowError} exception. To handle such exceptions, \texttt{PyGammaRAD} employs a dual-path arithmetic strategy. For the Clebsch-Gordan coefficient, a primary path first attempts evaluation using standard IEEE 754 double-precision floating-point arithmetic throughout, computing both the square-root prefactor and the alternating summation (see Eq.~(\ref{eq:cgc-1}), \ref{sec:cgc}) in native Python floats.  If an \texttt{OverflowError} is raised at any point, the entire calculation is automatically retried using the Python method \texttt{decimal.Decimal} with extended-precision arithmetic at 1000 significant digits.  In this fallback approach, the prefactor is evaluated via \texttt{decimal.Decimal.sqrt} and the alternating sum gets accumulated in a \texttt{Decimal} accumulator before conversion back to a float result upon return.  Wigner 3-$j$ symbols (see Eq.~(\ref{eq:w3j}), \ref{sec:3j}) and Gaunt coefficients (see Eq.~(\ref{eq:gaunt-2}), \ref{sec:gaunt}) are closely related to the Clebsch-Gordan coefficient, allowing for the same inherited treatment via standard double floating-point or extended precision, as appropriate.  For the Racah coefficient, which involves a product of four triangular delta factors each containing multiple factorials together with a similarly structured alternating sum (see Eq.~(\ref{eq:W1}), \ref{sec:racah}), the potential for numerical overflow is more acute; accordingly, \texttt{Decimal} arithmetic is employed unconditionally rather than as a fallback, ensuring that no precision is lost in intermediate calculations regardless of the magnitude of the angular momentum arguments.  Since the Wigner 6-$j$ symbol is obtained directly from the Racah coefficient through a simple phase factor (see Eq.~(\ref{eq:w6j}), \ref{sec:6j}), and the Wigner 9-$j$ symbol is decomposed into a sum of triple products of Wigner 6-$j$ symbols (see Eqs.~(\ref{eq:w9j-4}) and ~(\ref{eq:w9j-5}), \ref{sec:9j}), both quantities inherit the same extended-precision treatment.  It is worth noting that our tests of the Wigner 9-$j$ method encompass complete parity coverage: integral and half-integral cases, including mixed and large angular momentum entries, have been verified against published results and confirmed in the accompanying \texttt{Jupyter Notebooks}.

During development, various strategies were considered to avoid the pitfalls of overflow and underflow.  Although the methods described below are not actively used to return the results of any vector-coupling calculations described here, they have been left in the code for potential future use.  The triangular delta factor (see Eq.~(\ref{eq:cgc-2}), \ref{sec:cgc} and also \ref{sec:racah}) may be obtained independently via three different methods available within the \texttt{am\_formulae.Racah} class: direct evaluation \texttt{tri\_factor}; exponentiated logarithm \texttt{tri\_factor\_exp}; and logarithmic evaluation (to defer exponentiation) \texttt{tri\_factor\_log}.  The validity of the triangular delta-factor methods have been well-established through unit tests whereupon they were incorporated into the vector-coupling calculations described in \ref{sec:angmom} during earlier phases of development.  An integer square root method based on Newton's iterative method \cite{ypma:95}, implemented as a member function of the \texttt{am\_formulae.Newton} class as \texttt{isqrt}, is also provided as a utility for handling square roots of very large integers through refinement of an initial estimate based on the argument bit length until convergence with the correct integer square root is achieved, and has been tested against the native \texttt{math.sqrt} method using smaller integer arguments to ensure reproducibility.  Notebook actions and unit tests have also been developed to demonstrate the validity of the \texttt{am\_formulae.Newton.isqrt} method.  Finally, although we adopt the native \texttt{math.factorial} method and ensure arguments are interpreted as integers, we also performed checks against a user-defined recursive factorial method in addition to a $\Gamma$ function where $\Gamma(n+1) = n!$.  Consistency has also been demonstrated for these three methods, all of which are members of the \texttt{am\_formulae.Factorial} class.

\section{\label{sec:summary}Summary}

A set of calculators has been developed in Python to return precise $\gamma$-ray angular-distribution coefficients and statistical tensors that are useful in the interpretation of nuclear structure data.  These calculators have been benchmarked against the original canonical works of Yamazaki \cite{yamazaki:67} and Rose and Brink \cite{rose:67} for maximally-aligned nuclei, as well the tables of attenuation coefficients published by Der Mateosian and Sunyar \cite{mateosian:74} for partially-aligned nuclei.  Although the range of arguments presented in the tables of these original works \cite{yamazaki:67, mateosian:74, rose:67} may be useful in many cases, for most practical purposes in nuclear structure our calculators are no longer constrained by the size of these arguments and, e.g., calculations of coefficients involving nuclei populated to very high spin ($\gg 15\,\hbar$ cf. Ref.~\cite{yamazaki:67}) over a much wider range of multipolarities, magnetic substate projections, and for higher-rank tensors, may be readily evaluated.  A further feature of this program allows users to regenerate the original tables of Refs.~\cite{yamazaki:67, mateosian:74, rose:67} in convenient machine-readable CSV and JSON formats.  In particular, the defined vocabulary of the JSON schema provides implicit context regarding the nature and type of data held by the objects in their corresponding tables.  These feature-rich formats, together with the associated contextual information described in this project, enable straightforward feature engineering allowing for mapping of raw data into feature vectors for further applications where labeled $\gamma$-ray angular distribution data may be useful.  Extension of the tables to higher values of $J$, $L$, $m$, and $k$, as appropriate, can also be readily accomplished based on methods available within the \texttt{PyGammaRAD} library.

Given the essential role of angular momenta in these calculations, the software package is also loaded with a stand-alone general purpose vector-coupling calculator enabling exact analytical evaluation of Clebsch-Gordan, Racah, Wigner 3-$j$, 6-$j$, 9-$j$, and Gaunt coefficients.  These angular momentum coupling calculators have been verified against published data in various works on this topic \cite{stevenson:02, stone:80, tamura:70, wei:98, yukcu:25, johansson:16} and our results have also been checked using different proprietary \cite{Mathematica} and open \cite{SymPy} tools.  Notably, our vector-coupling methods can also accommodate and reproduce very large-integer couplings \cite{stevenson:02, johansson:16}.  This calculator combines the full apparatus of angular distribution calculations with a complete suite of vector-coupling algorithms that extends its use beyond the principal $\gamma$-ray angular-distribution focus to a broader range of quantum mechanical applications that involve the coupling of angular momenta, e.g., computation of molecular structure integrals based on Hartree-Fock-Roothaan methods \cite{roothaan:51}.  Finally, the small modular nature of \texttt{PyGammaRAD} allows it to be integrated into larger codebases where calculations of angular distributions or angular momentum coupling coefficients are needed.

\section*{Acknowledgments}

This work was performed under the auspices of the Department of Energy National Nuclear Security Administration by the Nuclear Science and Security Consortium under Award Number DE-NA0003996 at the University of California, Berkeley.  Additional support was received through the Lawrence Berkeley National Laboratory under Contract No. DE-AC02-05CH11231 and by the Los Alamos National Laboratory under Contract No. 89233218CNA000001.  We thank Dr. M. Cromaz for his valuable feedback and review of the manuscript.  One of the authors, Aaron, thanks Mr. William Bryn Hurst for his belief that still carries me through.

\appendix

\section{\label{sec:angmom} Angular momentum coupling calculator}

Although the development of a vector-coupling calculator is not a novel concept in itself, and indeed many such tools already exist that can satisfy this need, e.g., Refs.~\cite{tamura:70, wills:71, bretz:76, srinivasa:78, stone:80, lai:94, wei:98, stevenson:02, rasch:03, johansson:16, SymPy, Mathematica}, for convenience, the \texttt{PyGammaRAD} software package is also loaded with a complete angular momentum calculator that may serve a wide range of applications beyond the primary motivation of the work outlined in this paper.  

The computational treatment of vector-coupling coefficients has a long history 
rooted in Fortran implementations. The foundational program of Tamura 
\cite{tamura:70} provided subroutines for the Clebsch-Gordan and Racah 
coefficients based on direct evaluation of sum formulae; however, as noted by 
Wills \cite{wills:71}, this approach is susceptible to numerical overflow and 
round-off errors for large angular momenta in single precision, owing to the 
rapid growth of the intermediate factorials. Wills \cite{wills:71} addressed 
this for the Clebsch-Gordan coefficient by recasting the summation as a nested 
product, thereby avoiding repeated exponentiation of logarithmic sums. This 
improvement was subsequently extended to the Wigner 6-$j$ symbol by Bretz 
\cite{bretz:76}. Srinivasa Rao and Venkatesh \cite{srinivasa:78} then reformulated both the Clebsch-Gordan and Racah coefficients in terms of generalized hypergeometric functions of unit argument, specifically sets of $_3F_2(1)$ and $_4F_3(1)$ hypergeometric series, showing that this approach attains the same numerical accuracy as the Wills-Bretz method while achieving a further speed advantage of approximately 5--15\%. A distinct approach was taken by Stone and Wood \cite{stone:80}, who developed a Fortran package based on exact root-rational-fraction (RRF) arithmetic that addressed the limitations of large prime-factor storage which could cause integer overflow. Collectively, these Fortran implementations \cite{tamura:70, wills:71, 
bretz:76, srinivasa:78, stone:80} represent important milestones in the 
algorithmic development of vector-coupling coefficient computation, and their 
role in the lineage leading to modern implementations is recognized here. 
Computer algebra system approaches were subsequently developed by Lai 
\cite{lai:94}, who wrote programs in the Maple language capable of computing 
both algebraic formulae and numerically exact values of Wigner 3-$j$, 6-$j$, 
and 9-$j$ symbols; comparable symbolic capabilities are also available in 
Mathematica \cite{Mathematica}.  These tools, however, require proprietary 
software environments.  Stevenson \cite{stevenson:02} later addressed this accessibility concern with a suite of Java applets, openly available and capable of evaluating Clebsch-Gordan, Wigner 3-$j$, 6-$j$, and 9-$j$ symbols, and which remains accessible via GitHub \cite{COM:GitHub}. More recently, the third-party library \texttt{SymPy} \cite{SymPy} addressed this need through symbolic manipulation and arbitrary-precision evaluation within the open Python ecosystem.  \texttt{PyGammaRAD} represents a natural continuation of the open-source philosophy operating within the Python environment based on a conventional numerical computing approach, whilst broadening the scope to include the full angular distribution formalism.  An important contribution to the efficient handling of these coefficients in large-scale computations was made by Rasch and Yu \cite{rasch:03}, who presented storage schemes for precalculated Wigner 3-$j$, 6-$j$, and Gaunt coefficients that exploit the full set of symmetries these symbols possess, achieving retrieval speeds more than an order of magnitude faster than recursion-based evaluation.  A further important contribution in this area was developed by Johansson and Forss{\'e}n \cite{johansson:16} who demonstrated rapid precision evaluation of Wigner symbols using prime factorization and multiword integer arithmetic with an emphasis on couplings of very large angular momenta.  Fast evaluation of Gaunt and Clebsch-Gordan coefficients was also demonstrated by Xu \cite{xu:96, xu:97} based on a forward substitution lower-triangle linear system recurrence algorithm.  The \texttt{PyGammaRAD} implementation described here does not adopt a precalculation-and-storage strategy; instead, coefficients are evaluated on demand using the direct sum formulae described in the subsections below. This approach prioritizes simplicity, portability, and exact arithmetic over raw retrieval speed, and is appropriate for the angular distribution applications motivating this work where the number of required coefficient evaluations is modest.

In the strict sense, it is only the Clebsch-Gordan and Racah coefficients that are needed to evaluate the angular distribution coefficients presented earlier in Sect~\ref{sec:angdist}.  However, given their close relationship to the Wigner 3-$j$, 6-$j$, and 9-$j$ symbols typically used to describe coupling and recoupling schemes involving angular momentum in a quantum mechanical context, to broaden the utility of the \texttt{PyGammaRAD} software package, we provide methods to readily evaluate all aforementioned angular momentum coefficients and symbols.  To that end, we also include a method for direct evaluation of the Gaunt coefficient since it can be decomposed into a generalization of Wigner 3-$j$ symbols.  The algorithms upon which these quantities are based are defined explicitly below.

\subsection{\label{sec:cgc} Clebsch-Gordan coefficient}

The Clebsch-Gordan coefficient describes the coupling of angular momentum vectors $j_{1}$ and $j_{2}$, each with magnetic substate projections $m_{1}$ and $m_{2}$, respectively, to form a system of total angular momentum $j$ and corresponding projection $m$.  This coupling scheme may be expressed as
\begin{align}
  \label{eq:cgc-1}
  \notag
  \langle j_{1} m_{1} j_{2} m_{2} | j m \rangle &= \delta_{m, m_{1} + m_{2}} \Delta (j_{1}j_{2}j) \sqrt{(j+m)!(j-m)!(2j+1)} \\
  \notag
  &\times \sqrt{(j_{1}+m_{1})!(j_{1}-m_{1})!(j_{2}+m_{2})!(j_{2}-m_{2})!} \\
  &\times \sum\limits_{v = v_{\text{min}}}^{v_{\text{max}}} \left[ \frac{1}{(j-j_{2}+m_{1}+v)!(j-j_{1}-m_{2}+v)!} \right. \\
    \notag
    &\times \left. \frac{(-1)^{v}}{v!(j_{1}+j_{2}-j-v)!(j_{1}-m_{1}-v)!(j_{2}+m_{2}-v)!} \right],
\end{align}
where the triangular (or delta) factor is given by
\begin{equation}
  \label{eq:cgc-2}
  \Delta(j_{1}j_{2}j) = \sqrt{\frac{(j_{1}+j_{2}-j)!(j_{1}-j_{2}+j)!(-j_{1}+j_{2}+j)!}{(j_{1}+j_{2}+j+1)!}}.
\end{equation}
Only angular momentum vectors that satisfy the triangle inequalities condition result in the formation of a ($j_{1},j_{2},j$) triad arising from anti-parallel ($j_{1}-j_{2}$) and parallel ($j_{1}+j_{2}$) couplings:
\begin{equation}
  \label{eq:cgc-3}
  |j_{1} - j_{2} | \leq j \leq j_{1} + j_{2}.
\end{equation}
The summation over $v$ in Eq.~(\ref{eq:cgc-1}) runs over integers such that
\begin{align}
  \label{eq:cgc-4}
  \notag
  v_{\text{min}} &= \max (0, \max(-(j-j_{2}+m_{1}), -(j-j_{1}-m_{2})));\\
  v_{\text{max}} &= \min (j_{1}+j_{2}-j, \min(j_{1}-m_{1}, j_{2}+m_{2})).
\end{align}
The phase convention for the Clebsch-Gordan coefficients evaluated using Eq.~(\ref{eq:cgc-1}) follows that of Condon and Shortley \cite{condon:35}.  Additional conditions also apply regarding the magnetic quantum number projections: If $j_{i}$ is integral then so must be all corresponding $m_{i}$ projections, i.e., if $2j_{i} \bmod 2 = 0$ then $2m_{i} \bmod 2 = 0$.  Likewise, for half-integral $j_{i}$ and $m_{i}$ projections, if $2j_{i} \bmod 2 = 1$ then $2m_{i} \bmod 2 = 1$.  Also, for both integral and half-integral cases, each projection must satisfy the relation $|m_{i}| \leq j_{i}$.  Finally, the sum of the coupling projections must combine to give the total angular momentum projection, i.e., $m_{1} + m_{2} = m$.

\subsection{\label{sec:3j} Wigner 3-$j$ symbol}

The Wigner 3-$j$ symbol is based on the same sum formula used in Eq.~(\ref{eq:cgc-1}) to calculate the Clebsch-Gordan coefficient and the two quantities are related through the following expression
\begin{equation}
  \label{eq:w3j}
  \left( \begin{array}{ccc}
    j_{1} & j_{2} & j \\
    m_{1} & m_{2} & m \end{array} \right) = (-1)^{j+m+2j_{1}} \frac{1}{\sqrt{2j+1}} \langle j_{1} -m_{1} j_{2} -m_{2} | j m \rangle.
\end{equation}
The same conditions regarding the magnetic substate projections for the Clebsch-Gordan coefficient also apply here in the calculation of the Wigner 3-$j$ symbol, as does the triangle rule [Eq.~(\ref{eq:cgc-3})] for triad couplings.  Note that for projections $m_{1}$, $m_{2}$, and $m$ given in a Wigner 3-$j$ symbol, the projections must couple as $(-m_{1}) + (-m_{2}) = m$ (i.e., $m_{1}+m_{2}+m=0$).

\subsection{\label{sec:gaunt} Gaunt coefficient}

Gaunt coefficients arise in vector-coupling calculations where a triple product of spherical harmonics integrated over a solid angle is needed; this coefficient is included in \texttt{PyGammaRAD} for completeness.  Using the notation $\langle l_{2}m_{2}|l_{1}m_{1}|l_{3}m_{3}\rangle = Y^{l_{3}m_{3}}_{l_{1}m_{1}l_{2}m_{2}}$ \cite{yukcu:25}, we may then express the Gaunt coefficient as
\begin{align}
    \label{eq:gaunt-1}
    Y^{l_{3}m_{3}}_{l_{1}m_{1}l_{2}m_{2}} &=  \int\limits_{\Omega=0}^{\Omega=4\pi} Y_{l_{1},m_{1}}(\Omega) Y_{l_{2},m_{2}}(\Omega) Y_{l_{3},m_{3}}(\Omega) d\Omega\\
    \notag
    &=\int\limits_{\phi=0}^{\phi=2\pi} \int\limits_{\theta=0}^{\theta=\pi} Y_{l_{1},m_{1}}(\theta,\phi) Y_{l_{2},m_{2}}(\theta,\phi) Y_{l_{3},m_{3}}(\theta,\phi) \sin \theta d\theta d\phi,
\end{align}
where $Y_{l,m}(\Omega)$ is a spherical harmonic defined by orbital angular momentum and magnetic quantum numbers $l$ and $m$, respectively, and $d\Omega = \sin \theta d\theta d\phi$ is the infinitesimal element of solid angle subtended.  This Gaunt coefficient, as given by the integral of  Eq.~(\ref{eq:gaunt-1}), can be conveniently calculated as a generalization of two Wigner~3-$j$ symbols using the following expression
\begin{align}
    \label{eq:gaunt-2}
    G(l_{1}l_{2}l_{3},m_{1}m_{2}m_{3}) &= \sqrt{\frac{(2l_{1}+1)(2l_{2}+1)(2l_{3}+1)}{4\pi}} \\
    \notag
    & \times   
    \left( \begin{array}{ccc}
    l_{1} & l_{2} & l_{3} \\
    0 & 0 & 0 \end{array} \right)
    \left( \begin{array}{ccc}
    l_{1} & l_{2} & l_{3} \\
    m_{1} & m_{2} & m_{3} \end{array} \right).
\end{align}
Gaunt coefficients are governed by selection rules and have non-zero values only if $m_{1}+m_{2}+m_{3}=0$, the triangle condition $|l_{1} - l_{2}| \leq l_{3} \leq l_{1} + l_{2}$ is met, and $(l_{1} + l_{2} + l_{3}) \mod{2} = 0$.


\subsection{\label{sec:racah} Racah coefficient}

The Racah coefficient $W(j_{1}j_{2}Jj_{3}; J_{12}J_{23})$ involves the coupling of three angular momenta vectors.  Here, there are two associated coupling schemes.  In the first coupling scheme, $j_{1}$ couples with $j_{2}$ to form $J_{12}$, and then this resultant $J_{12}$ vector couples with $j_{3}$ to form total angular momentum $J$, i.e., $|((j_{1}j_{2})J_{12},j_{3})J\rangle$.  The other coupling scheme involves coupling $j_{2}$ with $j_{3}$ to give $J_{23}$, and $j_{1}$ next couples with $J_{23}$ to give the same resultant total angular momentum $J$, i.e., $|((j_{2}j_{3})J_{23},j_{1})J\rangle$.  Both coupling schemes result in complete orthonormal bases for the $(2j_{1} + 1)(2j_{2} + 1)(2j_{3} + 1)$ dimensional space, and the scalar product of these two angular momentum bases yields the Racah recoupling coefficient.  Explicitly, we may evaluate this coefficient as
\begin{align}
  \label{eq:W1}
  \notag
  W(j_{1} j_{2} J j_{3}; J_{12} J_{23}) &= \Delta(j_{1} j_{2} J_{12}) \Delta(J j_{3} J_{12}) \Delta(j_{1} J J_{23}) \Delta(j_{2} j_{3} J_{23}) \\
  &\times w(j_{1} j_{2} J j_{3}; J_{12} J_{23}),
\end{align}
where each of the four triangular factors may be evaluated analogously to Eq.~(\ref{eq:cgc-2}), subject to satisfaction of the triangle rule embodied by the inequality given in Eq.~(\ref{eq:cgc-3}), and
\begin{align}
  \label{eq:W2}
  \notag
  w(j_{1} j_{2} J j_{3}; J_{12} J_{23}) &\equiv \sum\limits_{z=z_{\text{min}}}^{z_{\text{max}}} \left[ \frac{(-1)^{z+\beta_{1}}(z+1)!}{(z-\alpha_{1})! (z-\alpha_{2})! (z-\alpha_{3})! (z-\alpha_{4})!} \right. \\
    &\times \left. \frac{1}{(\beta_{1}-z)! (\beta_{2}-z)! (\beta_{3}-z)!} \right],
\end{align}
where
\begin{align}
  \label{eq:W3}
  \notag
  \alpha_{1} &= j_{1} + j_{2} + J_{12}; \\
  \notag
  \alpha_{2} &= J + j_{3} + J_{12}; \\
  \notag
  \alpha_{3} &= j_{1} + J + J_{23}; \\
  \alpha_{4} &= j_{2} + j_{3} + J_{23},
\end{align}
and
\begin{align}
  \label{eq:W4}
  \notag
  \beta_{1} &= j_{1} + j_{2} + J + j_{3}; \\
  \notag
  \beta_{2} &= j_{1} + j_{3} + J_{12} + J_{23}; \\
  \beta_{3} &= j_{2} + J + J_{12} + J_{23}.
\end{align}
The summation over $z$ in Eq.~(\ref{eq:W2}) is finite and extends over limits given by
\begin{align}
  \label{eq:W5}
  \notag
  z_{\text{min}} &= \max(\alpha_{1},\alpha_{2},\alpha_{3},\alpha_{4});\\
  z_{\text{max}} &= \min(\beta_{1},\beta_{2},\beta_{3}).
\end{align}

\subsection{\label{sec:6j} Wigner 6-$j$ symbol}

Both the Racah coefficients and Wigner 6-$j$ symbols are a generalization of Clebsch-Gordan coefficients and Wigner 3-$j$ symbols that arise in the coupling of three angular momenta.  The Wigner 6-$j$ symbol and the Racah coefficient are related through a phase factor given by
\begin{equation}
  \label{eq:w6j}
  \left\{ \begin{array}{ccc}
    j_{1} & j_{2} & J_{12} \\
    j_{3} & J & J_{23} \end{array} \right\} = (-1)^{j_{1}+j_{2}+j_{3}+J} W(j_{1} j_{2} J j_{3}; J_{12} J_{23}).
\end{equation}
However, because of their higher number of symmetry properties, Wigner 6-$j$ 
symbols are often preferred in many algorithms, e.g., Ref.~\cite{rasch:03}, as a more efficient means of storing the recoupling coefficients\footnote{The 6-$j$ symbol possesses a total of 144 symmetries, corresponding to the $3! \cdot 4!$ row and column permutations of an associated $3 \times 4$ Regge array \cite{rasch:03}, compared with the 24 tetrahedral symmetries more commonly cited in standard texts. Rasch and Yu \cite{rasch:03} exploit this full set of symmetries to devise an efficient storage and retrieval scheme for precalculated 6-$j$ symbols, which can be indexed via an ordered one-dimensional array without redundancy.}.  Clearly, the usual constraint of the triangle rule given by Eq.~(\ref{eq:cgc-3}) also applies to angular momentum couplings involved in the evaluation of Wigner 6-$j$ symbols.  

In our implementation, the \texttt{am\_formulae.Racah} class serves as the computational engine for evaluation of both the Racah coefficient and the Wigner 6-$j$ symbol.  The \texttt{am\_methods.AngularMomentumCalculations.racah} function, which evaluates $W(j_{1}j_{2}Jj_{3};J_{12}J_{23})$ directly through \texttt{racah(j$_{\texttt{1}}$, j$_{\texttt{2}}$, J, j$_{\texttt{3}}$, J$_{\texttt{12}}$, J$_{\texttt{23}}$)}, calls the \texttt{am\_formulae.Racah} class with arguments reordered to match the internal convention, specifically as \texttt{Racah(j$_{\texttt{1}}$, j$_{\texttt{2}}$, J$_{\texttt{12}}$, j$_{\texttt{3}}$, J, J$_{\texttt{23}}$)}.  The \texttt{am\_methods.AngularMomentumCalculations.symb6j} function, which evaluates the Wigner 6-$j$ symbol
\begin{equation*}
  \left\{ \begin{array}{ccc}
    j_{1} & j_{2} & J_{12} \\
    j_{3} & J & J_{23} \end{array} \right\},
\end{equation*}
passes its arguments directly to the \texttt{am\_formulae.Racah} class as \texttt{Racah(j$_{\texttt{1}}$, j$_{\texttt{2}}$, J$_{\texttt{12}}$, j$_{\texttt{3}}$, J, J$_{\texttt{23}}$)}, with no reordering (i.e., in the same order as the function \texttt{symb6j(j$_{\texttt{1}}$, j$_{\texttt{2}}$, J$_{\texttt{12}}$, j$_{\texttt{3}}$, J, J$_{\texttt{23}}$)}), since the internal attribute ordering of the \texttt{am\_formulae.Racah} class was chosen to match the row-by-row layout of the Wigner 6-$j$ symbol.  The relationship between the two calling conventions follows directly from Eq.~(\ref{eq:w6j}): comparing the Wigner 6-$j$ symbol layout with the Racah coefficient arguments $W(j_{1}j_{2}Jj_{3}; J_{12}J_{23})$ shows that the top row of the Wigner 6-$j$ array $(j_{1}, j_{2}, J_{12})$ corresponds to $(j_{1}, j_{2}, J_{12})$ of the Racah coefficient, while the bottom row $(j_{3}, J, J_{23})$ corresponds to $(j_{3}, J, J_{23})$, confirming the positional rearrangement applied in the \texttt{racah} wrapper.

\subsection{\label{sec:9j} Wigner 9-$j$ symbol}

The Wigner 9-$j$ symbols are a generalization of Clebsch-Gordan coefficients, Wigner 3-$j$, and Wigner 6-$j$ symbols arising from coupling four angular momenta.  As with the Racah coefficient (or Wigner 6-$j$ symbol), the coupling of angular momenta can be achieved in different ways.  In this case, we have $j_{1} \otimes j_{2} = J_{12}$ and $j_{3} \otimes j_{4} = J_{34}$, before coupling these resultants $J_{12} \otimes J_{34}$ to give total angular momentum $J$, i.e., $|((j_{1},j_{2})J_{12},(j_{3},j_{4})J_{34})J\rangle$.  In the alternate coupling scheme, $j_{1} \otimes j_{3} = J_{13}$ and $j_{2} \otimes j_{4} = J_{24}$, before coupling $J_{13} \otimes J_{24} = J$, i.e., $|((j_{1},j_{3})J_{13},(j_{2},j_{4})J_{24})J\rangle$.  Again, both coupling schemes provide a complete orthonormal basis for the $(2j_{1}+1)(2j_{2}+1)(2j_{3}+1)(2j_{4}+1)$ dimensional space and the scalar product of the associated functions for these coupling schemes gives the Wigner 9-$j$ symbol.  

For computational purposes and to help simplify the coding task, it is convenient to decompose the Wigner 9-$j$ symbol into its triple product of constituent Wigner 6-$j$ symbols \cite{messiah:62,shore:68}:
\begin{align}
  \label{eq:w9j-1}
  \left\{ \begin{array}{ccc}
    j_{1} & j_{2} & J_{12} \\
    j_{3} & j_{4} & J_{34} \\
    J_{13} & J_{24} & J \end{array} \right\} &= \sum\limits_{x=x_{\text{min}}}^{x_{\text{max}}} (-1)^{2x} (2x + 1) \\
  \notag
  &\times \left\{ \begin{array}{ccc}
    j_{1} & j_{2} & J_{12} \\
    J_{34} & J & x \end{array} \right\}
  \left\{ \begin{array}{ccc}
    j_{3} & j_{4} & J_{34} \\
    j_{2} & x & J_{24} \end{array} \right\}
  \left\{ \begin{array}{ccc}
    J_{13} & J_{24} & J \\
    x & j_{1} & j_{3} \end{array} \right\}.
\end{align}
The summation in Eq.~(\ref{eq:w9j-1}) extends over all values of $x$ that satisfy the triangle inequality theorem in the corresponding triangular factors, such that no factorial has a negative argument within the limits given by
\begin{align}
  \label{eq:w9j-2}
  \notag
  x_{\text{min}} &= \max(|j_{1}-J|, |j_{2}-J_{34}|, |j_{3}-J_{24}|); \\
  x_{\text{max}} &= \min(j_{1}+J, j_{2}+J_{34}, j_{3}+J_{24}).
\end{align}
Due to its symmetry properties, any Wigner 9-$j$ symbol is invariant under reflection through its diagonals.  Therefore, the Wigner 9-$j$ symbol in Eq.~(\ref{eq:w9j-1}) has different equivalent representations, for example
\begin{align}
  \label{eq:w9j-3}
  \left\{ \begin{array}{ccc}
    j_{1} & j_{3} & J_{13} \\
    j_{2} & j_{4} & J_{24} \\
    J_{12} & J_{34} & J \end{array} \right\} &= \sum\limits_{x=x_{\text{min}}}^{x_{\text{max}}} (-1)^{2x} (2x + 1) \\
  \notag
  &\times \left\{ \begin{array}{ccc}
    j_{1} & j_{3} & J_{13} \\
    J_{24} & J & x \end{array} \right\}
  \left\{ \begin{array}{ccc}
    j_{2} & j_{4} & J_{24} \\
    j_{3} & x & J_{34} \end{array} \right\}
  \left\{ \begin{array}{ccc}
    J_{12} & J_{34} & J \\
    x & j_{1} & j_{2} \end{array} \right\},
\end{align}
whereupon the limits of $x$ are the same as those given in Eq.~(\ref{eq:w9j-2}).  

Computational methods based on formulations of alternative algorithms, rather than a summation over a triple product of Wigner 6-$j$ symbols, have also been proposed elsewhere. Lai \cite{lai:94} demonstrated symbolic and numerically exact computation of Wigner 9-$j$ symbols within the Maple computer algebra system, generating closed algebraic formulae for arbitrary angular momentum arguments. Wei \cite{wei:98} derived an algebraic expression for the Wigner 9-$j$ symbol as a twofold summation of products of binomial coefficients, with a recursive algorithm for evaluating those binomials that avoids the computation of factorials of integers entirely, thereby eliminating the primary source of overflow in large-angular-momentum calculations. 

The approach adopted in \texttt{PyGammaRAD} uses the triple Wigner 6-$j$ product decomposition of Eq.~(\ref{eq:w9j-1}), evaluated via the Racah sum formulae of the \texttt{am\_formulae.Racah} class [Eqs.~(\ref{eq:W1})--(\ref{eq:W5})].  In this implementation, the summation loop variable is stored internally as $h=2x$, an integer.  This ensures that the loop has a step size of two and always increments over integers rather than half-integers, with the phase factor $(-1)^{2x} \equiv (-1)^{h}$ and the degeneracy weight $(2x+1) \equiv (h+1)$ obtained directly from $h$.  This allows us to recast Eq.~(\ref{eq:w9j-1}) into the formal expression upon which the \texttt{PyGammaRAD} algorithm for the Wigner 9-$j$ symbol is computed:
\begin{align}
  \label{eq:w9j-4}
  \left\{ \begin{array}{ccc}
    j_{1} & j_{2} & J_{12} \\
    j_{3} & j_{4} & J_{34} \\
    J_{13} & J_{24} & J \end{array} \right\} &= \sum\limits_{h=h_{\text{min}}}^{h_{\text{max}}} (-1)^{h} (h + 1) \\
  \notag
  &\times \left\{ \begin{array}{ccc}
    j_{1} & j_{2} & J_{12} \\
    J_{34} & J & \frac{h}{2} \end{array} \right\}
  \left\{ \begin{array}{ccc}
    j_{3} & j_{4} & J_{34} \\
    j_{2} & \frac{h}{2} & J_{24} \end{array} \right\}
  \left\{ \begin{array}{ccc}
    J_{13} & J_{24} & J \\
    \frac{h}{2} & j_{1} & j_{3} \end{array} \right\},
\end{align}
with associated summation bounds given by
\begin{align}
  \label{eq:w9j-5}
  \notag
  h_{\text{min}} &= h_{\text{max}} \bmod 2; \\
  h_{\text{max}} &= 2 \min(j_{1}+J, j_{2}+J_{34}, j_{3}+J_{24}).
\end{align}
The lower bound $h_{\text{min}}$ in Eq.~(\ref{eq:w9j-5}) is set such that $h$ shares the same parity as $h_{\text{max}}$, ensuring $h$ increments in steps of two starting from either $h_{\text{min}} = 0$ (integral) or $h_{\text{min}} = 1$ (half-integral).  It is worth noting that the phase factor in Eq.~(\ref{eq:w9j-4}) will always be positive since $(-1)^{h}=1$ regardless of whether $h$ takes even or odd parity.  Each of the three Wigner 6-$j$ symbols in Eq.~(\ref{eq:w9j-4}) is evaluated by passing arguments directly to the \texttt{am\_formulae.Racah} class in the row-by-row layout as described in \ref{sec:6j}, with $h$ representing the intermediate angular momentum.




\bibliographystyle{elsarticle-num}
\bibliography{refs_pygammarad}







\end{document}